\documentclass[aps, prl, showpacs, showkeys, preprint, floatfix, superscriptaddress, nofootinbib, longbibliography]{revtex4-2}

\usepackage{graphicx}
\usepackage{amsfonts}
\usepackage{multirow}
\usepackage{color}

\def\vp{\varphi}

\def\half{\textstyle{\frac{1}{2}}}

\def\H{{\cal H}}

\def\p{\varphi}

\def\H{{\cal H}}

\def\ra{\rightarrow}
\def\tint{{\textstyle\int}}

\def\hp{{\hat\pi}}

\def\d{\partial}
\def\dag{\dagger}

\def\b{\begin{eqnarray*}}  
\def\e{\end{eqnarray*}}    
\def\bn{\begin{eqnarray}}  
\def\en{\end{eqnarray}}   

\def\<{\langle}
\def\>{\rangle}

\def\no{\nonumber}

\def\k{\kappa}

\def\hk{\hat{\kappa}}
\def\{{\lbrace}
\def\hv{\hat{\varphi}}

\def\}{\rbrace}
\begin{document}
\title{ Kinetic Factors in Affine Quantization and \\Their Role in Field Theory Monte Carlo}      
\author{Riccardo Fantoni}
\email{riccardo.fantoni@posta.istruzione.it}
\affiliation{Universit\`a di Trieste, Dipartimento di Fisica, \\
strada Costiera 11, 34151 Grignano (Trieste), Italy}   
\author{John R. Klauder}
\email{klauder@ufl.edu}
\affiliation{Department of Physics and Department of Mathematics \\
University of Florida,   
Gainesville, FL 32611-8440}

\date{\today}

\begin{abstract} Affine quantization, which is a parallel procedure with canonical quantization, needs to 
use its principal quantum operators, most simply $D=(PQ+QP)/2$ and $Q\neq0$, to represent appropriate kinetic factors, normally  $P^2$, which involve only one
canonical quantum operator. The need for this requirement stems from the quantization of selected problems that require affine quantization to achieve valid Monte Carlo results.
This task is resolved for introductory examples as well as examples  that 
involve scalar quantum field theories.
\end{abstract}

\maketitle

\section{Introduction}

In our previous articles, where some suitable Monte Carlo (MC) calculations  have been 
reported,  it was established that the quantum procedure called affine quantization (AQ) 
finds ``non-free'' results for the model  $\varphi^4_4$ \cite{Fantoni2020a,Fantoni2021a},  
while identical studies, which used canonical quantization (CQ), have only found ``free'' 
results, as if the coupling constant had been zero 
\cite{Freedman1982,Aizenman1981,Frohlich1982,Siefert2014,Aizenman2021}. After a careful 
comparison between the procedures of both CQ and AQ is presented, a detailed MC study of 
the model $(\varphi^2-\Phi^2)^2_4$ is presented. While the differences between AQ  and CQ 
for the first model are significant, the differences between AQ and CQ for the second 
model are much smaller, and a detailed study has found the reason why that could happen. 
Even if the AQ and CQ results for the second model are rather close, only one of those 
results can be physically correct.

A general effort to transform a variety of affine expressions opens up a variety of 
problems regarding their interaction terms and our present work was designed to do just 
that.

MC studies are greatly simplified by transforming affine variables
back into canonical variables, so the $\pi^2$ can join $(\sum_j d\varphi/dx_j)^2$
and imaginary time, to ensure a vast simplification of the MC
work. Such a transformation from affine to equivalent canonical
variables being required to achieve non-trivial results.

\section{Some Relations  Involving  the Quantum Operators $P$, $Q$, and $D$}

We need $[Q,P]=i\hbar 1\!\!1$, $F=F(Q)\neq0$, and we define $D=[PF+FP]/2$, so 
that $P^\dag F=PF$.\footnote{As affine quantization permits, the dilation operator, $D$, 
may take different forms, namely, $D =[P F(Q)+F(Q) P]/2$, for a variety of $F(Q)\neq0$ 
functions -- chosen such that $P^\dagger F(Q)=P F(Q)$ -- and which are of assistance in 
solving various problems.}
Then we examine 
 \bn &&\hskip-3em 2[F,D]= F(PF +FP) - (PF+FP) F \\
     &&\hskip0em  = FPF+FFP- PFF - FPF = FFP- PFF= [F^2,P] \:. \no \en
 This leads to $[F,D]= [F^2,P]/2 = i\hbar (F^2)'/2$, where the prime denotes a derivative with 
 respect to $Q$. As a familiar example, choose $F(Q)=Q$, 
 then $[Q,D] =[Q^2, P]/2 = i\hbar \,(Q^2)'/2=i\hbar \,Q$, analogues to the Lie algebra of the affine group \cite{ag}, and from which affine quantization got its name.

 \section{The Kinetic Factor in Hamiltonians}
  In simple problems, the most commonly chosen classical kinetic factor is $p^2$. In that realm we can choose $f(q)=1/g(q)\neq0$ ($ g(q)\neq0$ is added because $1/f(q)$ is very often used).
Now we define $d =p f(q)$
and we then recover $p^2$ from $d^2 g^2=d^2/f^2 =p^2$. Admittedly, this is utterly trivial. However,  when we quantize these variables to $P$, $D=(PF+FP)/2$, $F=F(Q)\neq0$ and $G = G(Q)=1/F(Q)\neq0$, difficulties can arise.

The quantum kinetic term (with $\hbar=1$) in affine variables is $DG^2D$.This expression, helped by $FP-PF =i \,F'$ and $ GP-PG =i \,G'$, leads to
\bn 4 DG^2D = (PF+FP) GG (PF+FP)=PP +FPGGPF+ FPGP +PGPF \no \\= PP + (PF+iF')GG(FP-iF') +
(PF+ iF')GP+ PG(FP-iF')  \no \\= 4PP  +2i(F'GP- PG F')  + F'GGF' = 4PP -2(F' G)' + (F')^2 G^2 \en
Restoring $\hbar$, it follows that 
\bn DG^2D = P^2+ (1/4)\hbar^2 [(F')^2 G^2 -2(F' G)']\:. \en

As a check on this expression, the example in which $F(Q)=Q$ and thus $G(Q)=1/Q$, leads to 
$ P^2 + (3/4)\hbar^2/Q^2$, which is the result previously found when $F(Q)=Q$.
There is every reason to accept this latter equation as the proper kinematical operator for the half-harmonic oscillator \cite{Klauder2020d,Gouba2020,Handy2021}.

\section{Application to Some Field Theory Examples}
\subsection{A straightforward example for $\vp^p_n$}
Regarding our field theory examples, our procedures will naturally encounter $\delta(0)$ divergences. A scaling procedure that eliminates such divergences will be introduced as sell as illustrated.
As our fist example we choose the classical canonical kinematic field $\pi(x)^2$, for which we 
choose the dilation field $\k(x)= \pi(x) \vp(x)$, with $\vp(x)\neq0$. The classical Hamiltonian 
in affine variables is
  \bn H_1 =\tint \{ \half[ \k(x)^2/\vp(x)^2 +(\nabla{\vp}(x))^2 +m^2\vp(x)^2] +g \,\vp(x)^p \}\;\d^s\!x \;,\en
  where $p=4,6,8,...$
  is the interaction power and $n=s+1$ is the number of spacetime dimensions. The advantage of this pair of variables is that $0<\vp(x)^{-2}<\infty$ which implies that $0<\vp(x)^p<\infty$,
 for all $p$, and thus the Hamiltonian does not experience any nonrenormalizability.
 
 Adopting the message from the half-harmonic oscillator, the affine quantum Hamiltonian 
 for this model is
  \bn \H_1=\tint \{\half [\hk(x)(\hv(x))^{-2}\hk(x) +(\nabla{\hv}(x))^2 + m^2 \hv(x)^2] + g\,
  \hv(x)^p \}\;d^s\!x \;, \en
  in which 
  \bn \hk(x)(\hv(x)^{-2})\hk(x)=\hp(x)^2+ (3/4)\hbar^2\delta(0)^{2s}/\vp(x)^2\:. \en
The origin of $\delta^s(0)=\infty$ is simply the fact that $[\hv(x),\hp(x)]=i\hbar \,\delta^s(0) 1\!\!1$. 

In a sense, this result is strange. For example, for a single classical variable $(pq)^2<\infty$ and 
$|QP-PQ|^2=\hbar^21\!\!1$. However, for a classical field $(\pi(x)\,\vp(x))^2<\infty$ while 
$|\hv(x)\,\hp(x)-\hp(x)\,\hv(x)|^2=\infty\,\hbar^2 1\!\!1$.
When approximated, as for an integration,  then $\hv(x)\ra \hv_k$ and $\hp(x)\ra \hp_k$, where instead of the continuum that $x$ represents,
$k$ identifies different points on a discrete lattice. This leads to $[\hv_k, \hp_k ]=i \hbar \,a^{-s} 1\!\!1$,
where  $a$ is a tiny spacial distance between neighboring lattice points. In preparation for our integration,  just as
every integral involves a continuum limit of an appropriate summation,  these expressions are used in Monte Carlo (MC) calculations which involve proper sums for their `integrals'. All of this is designed to provide a path integral quantization, and, when necessary,  their sums need to be regularized. 
In our case, the regularized version becomes appropriately   `scaled': specifically   $\vp_k\ra a^{-s/2}\vp_k$, $\pi_k\ra a^{-s/2}\pi_k$,
  $\k_k\ra a^{-s}\k_k$, $g\ra a^{s(p-2)/2}\,g$, and the regularized $d^s\!x \ra a^s$ may also be scaled as $a^s\ra a^{2s}$. 
  
  Using such scaling, in an AQ formulation with MC, has led to a  ``non-free'' result for the scalar field $\vp^4_4$ \cite{Fantoni2020a}. However, a CQ formulation with MC, along with analytic studies,  has led to a  ``free'' result  \cite{Freedman1982,Aizenman1981,Frohlich1982,Siefert2014,Aizenman2021,Fantoni2020a}.

  \subsection{A less common example using CQ and  AQ}
  With first using  CQ for the next example, our next classical Hamiltonian is given by
   \bn H_2= \tint\{\half[\pi(x)^2 +(\nabla{\vp}(x))^2 +m^2 \vp(x)^2] +g\, (\vp(x)^2- \Phi^2)^r \}
   \;d^s\!x \;, \en where the interaction power has been changed to $r=2,4,6,...$, and 
   $n=s+1$ is the same as before. This unusual  interaction term deserves a new dilation variable\footnote{Being able to change the dilation variable is an important feature of affine quantization.}, and in this section we choose $\k(x)=\pi(x)\,(\vp(x)^2-\Phi^2)$,
   where $(\vp(x)^2-\Phi^2)\neq0$. In this case, the classical Hamiltonian in affine variables becomes
  \bn  && H_3 =\tint \{\half [\k(x)^2/(\vp(x)^2-\Phi^2)^2 +(\nabla{\vp}(x))^2 +m^2 \vp(x)^2] \no \\ 
   &&\hskip10em + g\,(\vp(x)^2-\Phi^2)^r\} d^s\!x \;, \en
 In these variables, $0<(\vp(x)^2-\Phi^2)^{-2}<\infty $, which implies that $0<(\p(x)^2-\Phi^2)^r<\infty$, for all $r$, 
 thereby eliminating any nonrenormalizablity. 
 
 Next we find that the quantum Hamiltonian, using affine variables and Schr\"odinger's representation, is given by
  \bn && \H_3 =\tint \{ \half[\hk(x)(\vp(x)^2-\Phi^2)^{-2}\hk(x) +(\nabla{\vp}(x))^2 +m^2\vp(x)^2]
  \no \\ \label{eq:H3} &&\hskip10em  +g\,(\vp(x)^2-\Phi^2)^r \}\;d^s\!x\;, \en
  and this expression will become more useful after the kinetic term is fully analyzed.
 In order to obtain a valid quantum Hamiltonian for this model, we are first drawn back to Eq.~(3) in Sec.~2, 
 which reads $ DG^2D = P^2+ (1/4)\hbar^2 [(F')^2 G^2 -2(F' G)']$.
In the present case, temporally ignoring $(x)$ and still using Schr\" odinger's representation, $F=(\vp^2-\Phi^2)$ and $G=1/F$. It follows, that  $F'=2\vp$ and $G'=-2\vp/(\vp^2-\Phi^2)^2$. We also need 
$(F')^2 G^2=4\vp^2/(\vp^2-\Phi^2)^2$ and
$-2(F'G)'= -4/(\vp^2-\Phi^2)+ 8\vp^2/(\vp^2-\Phi^2)^2=4(\vp^2+\Phi^2)/(\vp^2-\Phi^2)^2$.   Hence, for this model, the kinematic factor is
 \bn &&\hk(x)(\vp(x)^2-\Phi^2)^{-2}\hk(x) \no \\
 &&\hskip3em =\hp(x)^2+\hbar^2\delta^{2s}(0)(2\vp(x)^2+\Phi^2)/(\vp(x)^2-\Phi^2)^2\;.\en
  As was the case in Sec.~3.1,  scaling can eliminate the $\delta^{2s}(0)$ factor by including  the additional scaling  factor $ \Phi^2\ra a^{-s}\Phi^2$, and changing the scaling 
  of $g$ to $g\ra a^{s(r-1)} g$.

\section{Lattice formulation of the field theory}
We used a lattice formulation of the AQ field theory stemming from the Hamiltonian of 
Eq. (\ref{eq:H3}) for $r=2$ and $s=3$ using the scaling 
$\vp\ra a^{-s/2}\vp, \Phi\ra a^{-s/2}\Phi, g\ra a^s g$ already employed in 
\cite{Fantoni2021,Fantoni2021a,Fantoni2021b}. The 
theory considers a real scalar field $\vp$ taking the value $\vp_k$ on each site of a 
periodic, hypercubic, $n$-dimensional lattice of lattice spacing $a$, our ultraviolet 
cutoff, and periodicity $L=Na$. Using the usual classical expression $\pi=d\vp/dt$, where 
$t$ is imaginary time, for the momentum field, the affine action, $S=\int \H_3\,dx_0$, 
with $x_0=ct$ where $c$ is the speed of light constant, is then approximated on the 
lattice by
\bn \label{eq:scaled-affine-action} \nonumber
S[\vp]/a^{n-s}&\approx&\half\left\{\sum_{k,\mu}a^{-2}(\vp_k-\vp_{k+e_\mu})^2 
+m^2\sum_{k}\vp_k^2\right\}+\sum_{k}g\,(\vp_k^2-\Phi^2)^2\\
&&+{\textstyle\frac{1}{2}\sum_{k}}\hbar^2{\displaystyle\frac{2\vp_k^2+\Phi^2}{(\vp_k^2-\Phi^2)^2}},
\en
where $e_\mu$ is $1$ in the $+\mu$ direction and $0$ else. This is known as the 
{\sl primitive approximation} for the action and could be improved in various ways 
\cite{Ceperley1995}. For the CQ field theory the last term in 
(\ref{eq:scaled-affine-action}), proportional to $\hbar^2$ should be dropped.

In this work we are interested in reaching the continuum limit by taking $Na$ fixed 
and letting $N\to\infty$ at fixed volume $L^s$ and absolute temperature $T=1/k_BL$ with 
$k_B$ the Boltzmann's constant. We will always work in natural units $c=\hbar=k_B=1$.

\section{PIMC results}
We performed path integral MC \cite{Metropolis,Kalos-Whitlock,Ceperley1995,Fantoni12d} 
calculation for the AQ field theory described by Eq. (\ref{eq:scaled-affine-action}) for 
$n=3+1$ and $\Phi=1$, and compared it with the corresponding CQ field theory. In 
particular we calculated the renormalized coupling constant $g_R$ (which must be 
non-negative due to Lebowitz inequality) and mass $m_R$ defined 
in Eqs. (4.3) and (4.5) of \cite{Fantoni2020} respectively. This will allow us to explore 
the behavior of the renormalized system, for a given set of parameters $m,g$, as a 
function of $N$ at fixed volume and temperature.

\begin{figure}[htbp]
\begin{center}
\includegraphics[width=8cm]{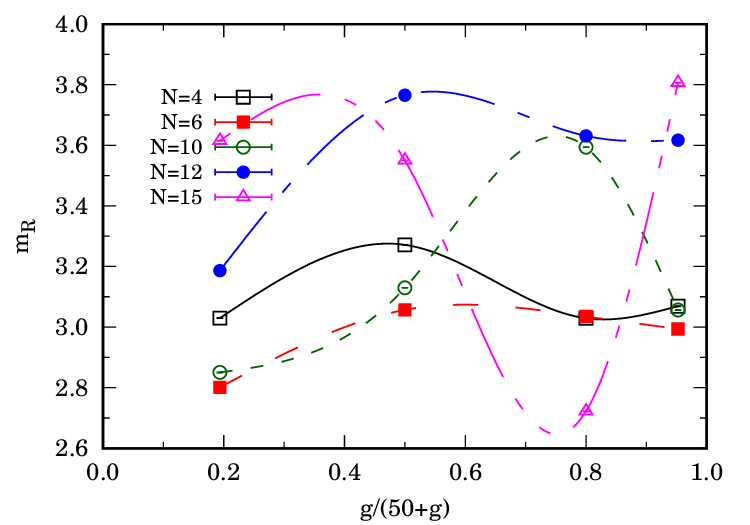}\\
\includegraphics[width=8cm]{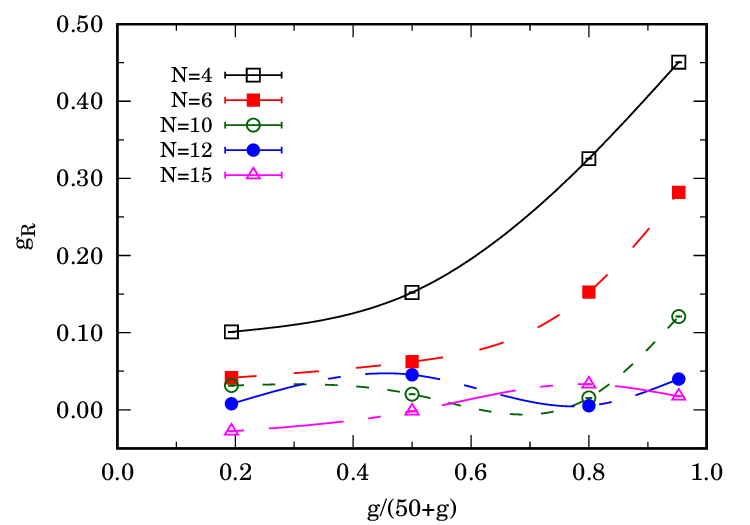}\\
\includegraphics[width=8cm]{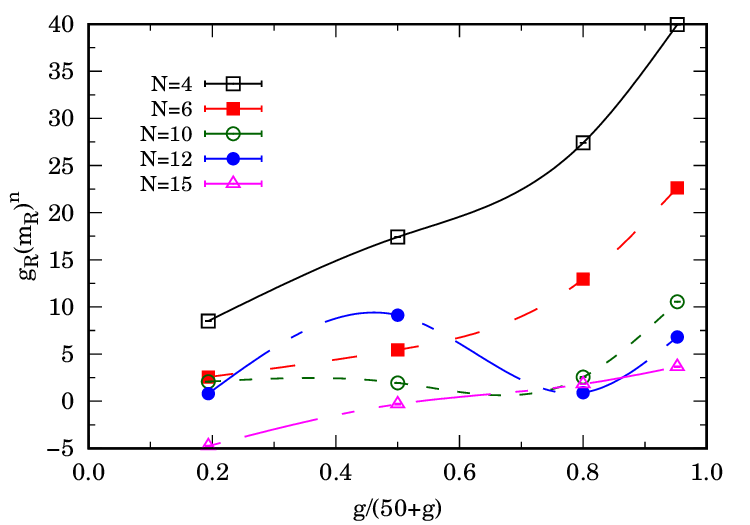}
\end{center}
\caption{AQ case. We show the renormalized mass $m_R\approx 3$ (top panel), the 
renormalized coupling constants $g_R$ (central panel), and $g_Rm_R^n$ (bottom panel) 
for various values of the bare coupling constant $g$ at decreasing 
values of the lattice spacing $a=1/N$ ($N\to\infty$ continuum limit) for the 
{\sl scaled affine} covariant euclidean scalar field theory described by 
the lattice action of Eq. (\ref{eq:scaled-affine-action}) for $n=3+1$ and $\Phi=1$. The 
lines connecting the simulation points are just a guide for the eye. The lack of errorbars in the data presented is justified by
the fact that the errors are dominated not from the statistical ones
but rather from the ones due to the adjustments in the bare mass
required by the trial and error procedure suggested by Freedman
et. al. \cite{Freedman1982}. This error is very hard to be estimated.}
\label{fig:123}
\end{figure}

\begin{figure}[htbp]
\begin{center}
\includegraphics[width=8cm]{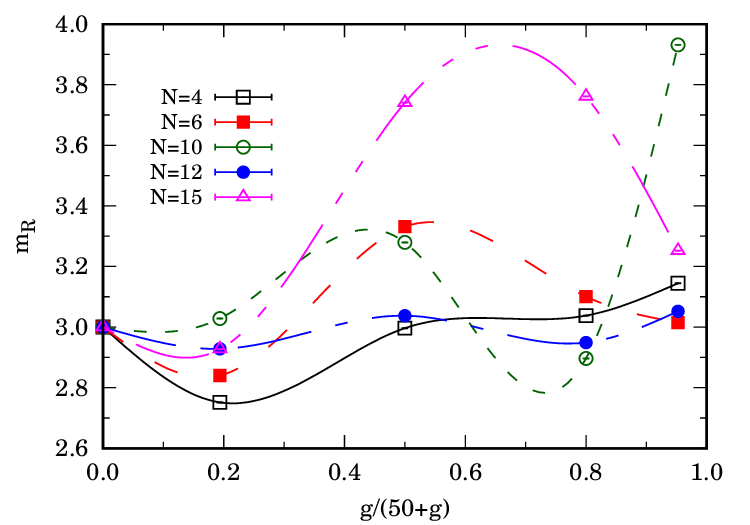}\\
\includegraphics[width=8cm]{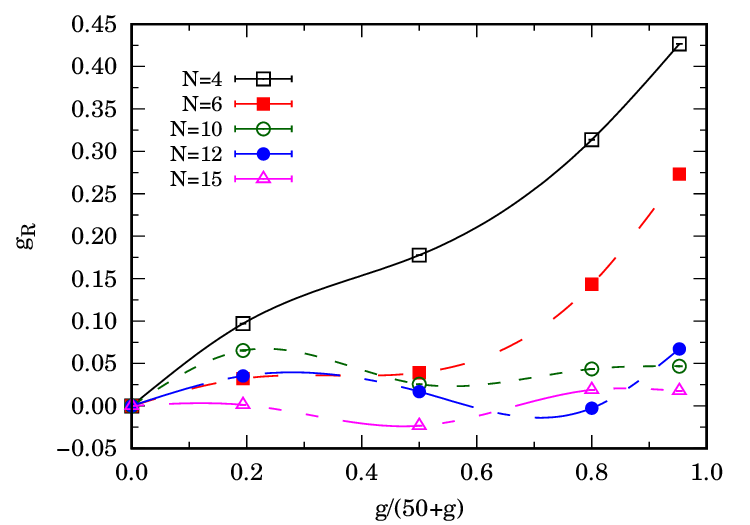}\\
\includegraphics[width=8cm]{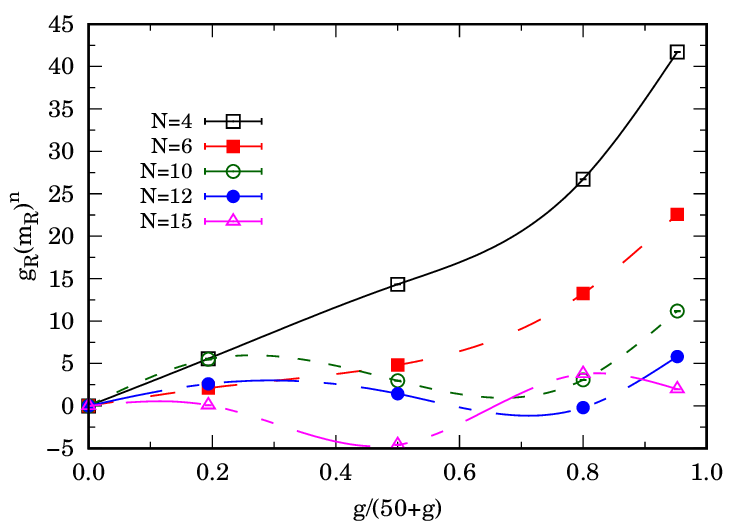}
\end{center}
\caption{CQ case. We show the renormalized mass $m_R\approx 3$ (top panel), the 
renormalized coupling constants $g_R$ (central panel), and $g_Rm_R^n$ (bottom panel) 
for various values of the bare coupling constant $g$ at decreasing 
values of the lattice spacing $a=1/N$ ($N\to\infty$ continuum limit) for the 
{\sl canonical} covariant euclidean scalar field theory described by 
the lattice action of Eq. (\ref{eq:scaled-affine-action}) without the last term 
proportional to $\hbar^2$, for $n=3+1$ and $\Phi=1$. The lines connecting the simulation 
points are just a guide for the eye. The lack of errorbars in the data presented is justified by
the fact that the errors are dominated not from the statistical ones
but rather from the ones due to the adjustments in the bare mass
required by the trial and error procedure suggested by Freedman
et. al. \cite{Freedman1982}. This error is very hard to be estimated.}
\label{fig:123-c}
\end{figure}

Following Freedman et al. \cite{Freedman1982}, for each $N$ and $g$, we adjusted the 
bare mass $m$ in such a way to maintain the renormalized mass approximately constant 
$m_R\approx 3$ to within a few percent (in all cases less than $25\%$). Differently 
from our previous study \cite{Fantoni2020a} with the unscaled version of the affine 
field theory we did not need to choose complex $m$ in order to fulfill this constraint, 
as shown in Table \ref{tab:m}. In fact our present CQ model can be obtained from the 
$\vp^4_4$ model studied in \cite{Fantoni2020a} by changing 
$m^2\to m^2-4g\Phi^2\equiv M^2$ which will become negative for $g$ big enough. From the 
Table we can see how for the chosen cases $m^2/4g\sim\Phi^2$, meaning that the minima 
$\vp_\pm=\pm \sqrt{-M^2/4g}$ of the potential profile
${\cal V}[\phi]=m^2\vp^2/2+g\,(\vp^2-\Phi^2)^2$ are far from $\pm\Phi$, where the 
effective potential term, $(2\vp^2+\Phi^2)/2(\vp^2-\Phi^2)^2$, stemming from the kinetic 
part of the action (the last term in Eq. (\ref{eq:scaled-affine-action}) proportional to 
$\hbar^2$) diverges. As a consequence CQ will be very similar to AQ. Which means that the 
required bare masses to reach a given renormalized mass in the two cases are very close.
\begin{table}[htbp]
\caption{Choice of the bare mass $m$ in the simulations for CQ and AQ cases. Also shown 
is $M^2=m^2-4g\Phi^2$ and $m^2/4g\Phi^2$.} 
\label{tab:m}
\vspace{.5cm}
{\footnotesize \begin{tabular}{||c|c|ccc||ccc||}
\hline
\hline
&&
\multicolumn{3}{|c||}{CQ}&
\multicolumn{3}{|c||}{AQ}\\
\hline
$N$ & $g$ & $m$ & $M^2$ & $m^2/4g\Phi^2$ & $m$ & $M^2$ & $m^2/4g\Phi^2$\\
\hline 
\multirow{4}{*}{4} & 12 & 7.00 & 1.000 & 1.021 & 6.65 & $-3.777$ & 0.921 \\
 & 50 & 13.70 & $-12.31$ & 0.938 & 13.55 & $-16.397$ & 0.918 \\
 & 200 & 27.20 & $-60.16$ & 0.925 & 27.10& $-65.590$ & 0.918 \\
 & 1000 & 61.25 & $-248.438$ & 0.938 & 61.20 & $-254.56$ & 0.936 \\
\hline
\multirow{4}{*}{6} & 12 & 7.20 & 3.840 & 1.080 & 6.80 & $-1.76$ & 0.963 \\
 & 50 & 14.00 & $-4.000$ & 0.980 & 13.75 & $-10.937$ & 0.945 \\
 & 200 & 27.50 & $-43.750$ & 0.945& 27.40 & $-49.240$ & 0.938 \\
 & 1000 & 61.57 & $-209.135$ & 0.948 & 61.53 & $-214.059$ & 0.946 \\
\hline 
\multirow{4}{*}{10} & 12 & 7.40 & 6.760 & 1.141 & 7.00 & 1.000 & 1.021 \\
 & 50 & 14.20 & 1.640 & 1.008 & 14.00 & $-4.000$ & 0.980 \\
 & 200 & 27.80 & $-27.160$ & 0.960 & 27.80 & $-27.160$ & 0.960 \\
 & 1000 & 62.10 & $-143.590$ & 0.964 & 62.00 & $-156.000$ & 0.961 \\
\hline 
\multirow{4}{*}{12} & 12 & 7.40 & 6.760 & 1.141 & 7.30 & 5.29 & 1.110 \\
 & 50 & 14.20 & 1.640 & 1.008 & 14.20 & 1.640 & 1.008 \\
 & 200 & 27.90 & $-21.590$ & 0.973 & 27.90 & $-21.590$ & 0.973 \\
 & 1000 & 62.20 & $-131.160$ & 0.936 & 62.20 & $-131.160$ & 0.936 \\
\hline 
\multirow{4}{*}{15} & 12 & 7.40 & 6.760 & 1.141 & 7.40 & 6.760 & 1.141 \\
 & 50 & 14.40 & 7.36 & 1.037 & 14.20 & 1.640 & 1.008 \\
 & 200 & 28.10 & $-10.390$ & 0.987 & 27.90 & $-21.590$ & 0.973 \\
 & 1000 & 62.40 & $-106.240$ & 0.973 & 62.40 & $-106.240$ & 0.973 \\
\hline
\hline
\end{tabular}}
\end{table}
Then we measured the renormalized coupling constant $g_R$ defined in 
\cite{Fantoni2020,Fantoni2020a} for various values of the bare coupling constant $g$ at a 
given small value of the lattice spacing $a=1/N$ (this corresponds to choosing a fixed 
absolute temperature $k_BT=1$ and a fixed volume $L^3=1$) as already explained for 
example in Refs. \cite{Fantoni2020,Fantoni2020a}. With $Na$ and $m_R$ fixed, 
as $a$ was made smaller, whatever change we found in $g_Rm_R^n$ as a function of $g$ 
could only be due to the change in $a$. We generally found that a depression in $m_R$ 
produced an elevation in the corresponding value of $g_R$ and viceversa. The results 
are shown in Fig. \ref{fig:123} for the scaled affine action (AQ case) 
(\ref{eq:scaled-affine-action}), where, following Freedman et al. \cite{Freedman1982} we 
decided to compress the range of $g$ for display, by choosing the horizontal axis to be 
$g/(50+g)$. For comparison we also show in Fig. \ref{fig:123-c} the results for canonical 
quantized action (CQ case) which is given by Eq. (\ref{eq:scaled-affine-action}) without 
the last term proportional to $\hbar^2$. The constraint $m_R\approx 3$ was not easy to 
implement since for each $N$ and $g$ we had to run the simulation several (5-10) times 
with different values of the bare mass $m$ in order to determine the value which would 
satisfy the constraint $m_R\approx 3$. In our simulations we always used $3\times 10^7$ 
MC sweeps (where one sweep moves all the $N^n$ field points which took about one week of 
computer time for the $N=15$ case). We estimated 
that it took roughly $10\%$ of each run in order to reach equilibrium from the 
arbitrarily chosen initial field configuration, for each set of parameters.


As we can see from our Figures, the renormalized coupling constant $g_R(m_R)^4$ of the 
scaled affine version (AQ of Fig. \ref{fig:123}) behaves very similarly to the one of the 
canonical version (CQ of Fig. \ref{fig:123-c}) going towards the continuum limit, taken 
at fixed volume and temperature, when the ultraviolet cutoff is gradually removed ($Na=1$ 
and $N\to\infty$). The only difference is at $g=50-100$ where in the AQ case the $N=12$ 
results for the renormalized coupling fall above the ones for $N=10$, unlike what happens 
in the CQ case. Note that for the CQ case the results at $N=12,15$ are new, since 
Freedman et al. \cite{Freedman1982} and ourselves \cite{Fantoni2020a} only previously 
studied up to $N=10$ discretization  points.

During our simulations we kept under control also the vacuum expectation value of the 
field which is not diverging going towards the continuum limit, like what was happening 
in \cite{Fantoni2021} but not in \cite{Fantoni2020b}. Choosing the initial configuration 
with $\vp=0$ at all lattice points, when $M^2$ is not too negative the symmetry 
$\vp\to-\vp$ is not broken and we find $\langle\vp\rangle\sim 0$.

We also studied the behavior of the AQ case when choosing a much lower renormalized mass 
$m_R\sim 1/10$. In this case the necessary bare mass is such that $m^2/4g\ll\Phi^2$, at 
all studied values of the bare coupling $g=12,50,200,1000$. In 
particular the potential profile ${\cal V}$ becomes a symmetric double well with the two 
minima, at $\vp_\pm$, near the two repulsive spikes localized at $\vp=\pm\Phi$ and 
forbidding paths to access the minima of the double well. 
\footnote{The case when the classical minima of the potential and the extra
spikes in the potential of the affine Hamiltonian are close together
has been already studied in several of our previously published
papers \cite{Fantoni2020,Fantoni2020a,Fantoni2021a,Fantoni2021b,Fantoni2022}.} 
In this case we found that the 
paths tend to be very localized just outside of the forbidden region due to the repulsive 
spikes. As a consequence we found $g_R\sim 2$ for all $N$. So in this case AQ is very 
different from CQ and the bare masses necessary to reach the same renormalized mass are 
very different. Note that when $M^2>0$ the two repulsive spikes do not forbid the path 
from sitting at the minimum of the potential profile at $\vp=0$ and as a consequence AQ 
and CQ are very similar. Note also that in the limit $\Phi\to 0$ the situation is 
inverted and for $m^2$ positive, AQ is very different from CQ, whereas for $m^2$ 
negative, AQ is very similar to CQ.

\section{Conclusions}

We studied through path integral Monte Carlo a plausible kinetic factor in affine 
quantization (AQ) of a scalar covariant euclidean field theory of mass $m$ subject to a 
potential energy of the form $g(\vp^2-\Phi^2)^2$ in $3+1$ space-time dimensions, which is 
known to suffer from asymptotic freedom in the continuum limit when it is quantized 
through canonical quantization (CQ). This kinetic factor reduces to the usual one 
previously introduced in 
\cite{Fantoni2020,Fantoni2020a,Fantoni2020b,Fantoni2021,Fantoni2021a,Fantoni2021b} in the 
limit $\Phi\to0$, apart from the multiplicative coefficient. Moreover its behavior is 
similar to the one found in the $\Phi\to0$ limit in the sense that it gives rise to an 
additive effective potential term which diverges in a neighborhood of the minima in the 
potential therefore producing a forbidden region for the field paths exactly where it 
would naturally sit in a canonical quantization framework. This exclusion of the field 
path from the minima of the potential renders the affine quantization version of the 
field theory asymptotically non-free in the continuum limit.

Our numerical results clearly show how the two field theories obtained through CQ and AQ 
behave very differently whenever $m^2/4g\ll\Phi^2$. Otherwise they are very similar.

\bibliography{killer}


\end{document}